\documentstyle[12pt]{article}

\def\+{{+\!\!\!+}}
\def\pp{\mbox{\tiny${}_{\stackrel\+ =}$}}

\def\d{\partial}

\def\th{\theta}

\def\P{\Phi}

\def\r{\rho}

\def\L{\Lambda}

\def\s{\sigma}
\def\p{\psi}

\def\e{\varepsilon}

\def\pmb#1{\setbox0=\hbox{#1}%
\kern.0em\copy0\kern-\wd0
\kern-.04em\copy0\kern-\wd0
\kern.08em\copy0\kern-\wd0
\kern-.04em\raise.0433em\box0 }         

\def\ihalf{{\textstyle{i \over 2}}}

\newcommand{\nc}{\newcommand}
\nc{\beq}{\begin{equation}}
\nc{\eeq}[1]{\label{#1}\end{equation}}
\nc{\ber}{\begin{eqnarray}}
\nc{\eer}[1]{\label{#1}\end{eqnarray}}
\nc{\pek}[1]{\cite{#1}}
\nc{\enr}[1]{(\ref{#1})}
\nc{\kal}[1]{{\cal{#1}}}
\nc{\dott}{\;\cdot\;}

\begin{document}
\newcommand{\inv}[1]{{#1}^{-1}} 
\renewcommand{\theequation}{\thesection.\arabic{equation}}
\newcommand{\be}{\begin{equation}}
\newcommand{\ee}{\end{equation}}
\newcommand{\bea}{\begin{eqnarray}}
\newcommand{\eea}{\end{eqnarray}}
\newcommand{\re}[1]{(\ref{#1})}
\newcommand{\qv}{\quad ,}
\newcommand{\qp}{\quad .}
\begin{center}

                                \hfill    USITP-00-03\\
                                \hfill    hep-th/0004061\\

\vskip .3in \noindent

\vskip .1in

{\large \bf {Boundary Conditions, Supersymmetry and $A$-field Coupling
for an Open String in a $B$-field Background}}
\vskip .2in

{\bf Parviz Haggi-Mani}\footnote{parviz@physto.se},
 {\bf Ulf Lindstr\"om}\footnote{ul@physto.se, supported by the Swedish
 Natural Science Research Council}
 and {\bf Maxim Zabzine}\footnote{zabzin@physto.se}  \\

\vskip .15in

\vskip .15in

\vskip .15in

 {\em Institute of Theoretical Physics,
University of Stockholm \\
Box 6730,
S-113 85 Stockholm SWEDEN} \\
\bigskip

\vskip .15in

\vskip .1in
\end{center}
\vskip .4in
\begin{center} {\bf ABSTRACT }
\end{center}
\begin{quotation}\noindent
We discuss the non-linear sigma model representing a NSR open
string in a curved background with non-zero $B_{\mu\nu}$-field. With
this coupling the theory is not automatically supersymmetric,
due to boundary contributions. When $B=0$ supersymmetry is
ensured by the conditions that follow as the boundary
contribution to the field equations. We show that inclusion of
a particular boundary term restores this state of affairs also
in the presence of a $B$-field. The boundary conditions derived
from the field equations in this case agree with those
that have been proposed for constant $B$-field. A
coupling to a boundary $A_\mu$-field will modify both the
boundary conditions and affect the supersymmetry. It is shown
that there is an $A$-coupling with non-standard fermionic part
that respects both the supersymmetry and the shift symmetry (in
the $B$ and $A$ fields), modulo the (modified) boundary
conditions.

\end{quotation}
\vfill
\eject

\section{Introduction}

Lately, interest in relevance of noncommutative geometry for string
 theory \cite{Connes:1998cr} has led to investigations of open
 strings propogating in a (constant) two-form background \cite{Chu:1998qz, 
 Seiberg:1999vs}.
 In this context generalization of the periodic (antiperiodic) 
 boundary condition for open string fermions (and the bosonic
 counterpart) have been discussed
  \cite{Schomerus:1999ug, Chu:1999gi, Seiberg:1999vs}, but not derived from
 an action. 
In this paper we study the (globally) supersymmetric sigma model action
representing an open NSR 
string in a curved background with non-zero $B$-field. In the most
general setting, both the metric and the $B$-field are taken to be
arbitrary background fields.  With this coupling the theory is not
automatically supersymmetric, due to boundary contributions. When $B=0$,
however,  supersymmetry is ensured by the conditions that follow as the
boundary contribution to the field equations. We show that inclusion of a
particular boundary term restores this state of affairs also in the 
general case. The boundary conditions derived from the field equations in
this case agree with those that have been discussed for constant $B$-field.
We further consider the
coupling to a boundary $A_\mu$-field. Such a coupling will modify the
boundary conditions and affect the supersymmetry, but is needed for
invariance under the combined gauge transformation of the $B$-field and
shift of the $A$-field. It turns out that there is a
$A$-coupling with non-standard fermionic part that respects both the
supersymmetry and the shift symmetry  modulo
the (modified) boundary conditions. A remakable feature of this
$A$-coupling on the boundary is that, whereas it is not supersymmetric by
itself, the boundary conditions nevertheless ensures invariance under
supersymmetry (of the whole action).

 We emphasize that our attitude is to take seriously the
boundary 
 conditions derived from the total action, including background
 $B$ field and boundary $A$-field, and try to reconcile them both with 
 supersymmetry (and shift symmetry).

 The paper is organized as follows:
Section $2.1$  deals with the known case of a constant
$B$-field and a flat Minkowski metric. Here we rederive the known boundary
conditions from an action by adding a boundary term to the standard
Lagrangian. We also prepare the ground for the generalisation to a
non-constant metric and $B$-field presented in Section $2.2$.
 In Section $3$ we give the boundary coupling of the $A$-field, the
corresponding boundary conditions for the whole action and the proof
of boundary supersymmetry. In Section $4$, for completeness, we
sketch the covariant quanization of the model with constant background
fields,
 discuss the question of space-time symmetry, exhibit the 
 breaking of the Lorentz
group and construct the vertex operator for a massless boson. 
 Section $5$ contains our comments and conclusions.

\section{Construction of the action}

\subsection{Constant
$B$-field}

In this subsection we present the construction of the globally
supersymmetric world-sheet action for the open string in Minkowski space in a
constant $B$-field. (The discussion in the literature
 has been rather inconclusive \cite{Chu:1998qz}-\cite{Chu:1999gi}.)
 To display the ideas, the presentation
will be very explicit. Exactly the same
logic will be applied to the non-constant case in the next
subsection.

 Many questions can be dealt with
 without explicit knowledge of the correct action. A constant $B$-field
 will not modify the bulk physics, (the equations of motion
 and the stress tensor,e.g.),
but only the boundary conditions. From the bosonic
 theory we know the correct boundary condition for the coordinate $X^\mu$:
\be
\label{bosbound}
[E_{\nu\mu}\d_\+ X^\nu-E_{\mu\nu}\d_= X^\nu]_{|_{\sigma=0,\pi}}=0 ,
\ee
 where
$E_{\mu\nu}=\eta_{\mu\nu} + B_{\mu\nu}$.
 Using the supersymmetry
transformations (see Appendix for notation)
\bea
\label{susytransf}
\delta X^\mu&=& -\epsilon^{+}\psi^\mu_{+} - \epsilon^{-} \psi^\mu_{-}, \\
\delta \psi^\mu_{+}&=&-i\epsilon^{+} \d_\+ X^\mu ,\\ 
\delta \psi^\mu_{-}&=&-i\epsilon^{-} \d_= X^\mu,
\eea
 we find (\ref{bosbound}) to be the boundary supersymmetry transformation of
an expression involving the fermions 
\be
\label{varfer}
\eta
\left( \eta_{\mu\nu} X'^{\nu}
-B_{\mu\nu}\dot{X}^\nu\right)|_{\sigma=0,\pi}
= \frac{i}{2}\delta\left( E_{\nu\mu}\psi_{+}^\nu \mp
 E_{\mu\nu}\psi_{-}^\nu \right )|_{\sigma=0,\pi},
\ee
 where $\eta \equiv\epsilon^{+} = \pm \epsilon^{-}$, $\eta_{\mu\nu}$ is the
Minkowski metric. For the model to be
supersymmetric, we thus have to require the expression on the right hand
side to vanish. Our task is then to construct an action which gives rise
 to both (\ref{bosbound}) and the fermionic boundary condition implicit
in (\ref{varfer}).
 We start from the standard superfield action 
\be
\label{sup}
S = \frac{1}{4\pi\alpha'}
\int d^2\xi\,d^2\theta\,\,D_{+} \Phi^\mu D_{-} \Phi^\nu E_{\mu\nu},
\ee
 with constant
  $E_{\mu\nu}$ and manifest bulk supersymmetry. 
 The component action that results from (\ref{sup}) is
$$S= -\frac{1}{4\pi \alpha'}\int d^2\xi\, (
\partial_\alpha X_\mu \partial^\alpha X^\mu
+i\overline\psi^\mu \rho^\alpha \partial_\alpha \psi_\mu+$$
\be
\label{compaction}
+\epsilon^{\alpha \beta}B_{\mu \nu}\partial_\alpha X^\mu \partial_\beta
 X^\nu
+i\epsilon^{\alpha \beta}B_{\mu \nu}\overline\psi^\mu \rho_\alpha
\partial_\beta \psi^\nu )
\ee
 where we have kept all boundary terms. Varying the action we find
 the usual field equations and the boundary contributions, which should
vanish. We find that we must require
\be
\label{fer-bc}
 \left( \delta X^\mu
 [\d_= X^\nu E_{\mu\nu} - \d_\+ X^\nu E_{\mu\nu} ] +
 i  [\psi^\mu_{-}\delta \psi^\nu_{-} E_{\nu \mu} -
 \psi^\mu_+\delta \psi^\nu_+ E_{\mu \nu} ]
 \right )_{|_{\sigma=0,\pi}}=0 .
\ee
 An additional condition follows from 
 the the supersymmetry variation of the
 action (\ref{compaction}), namely
\be
\label{variation}
\eta[\d_\+ X^\mu  \psi^\nu_{-}  E _{\mu \nu} \mp
   \psi^\mu_+ 
 \d_=X^\nu E_{\mu\nu}]_{|_{\sigma=0,\pi}}=0 .
\ee
 Correct 
 boundary conditions for $X^\mu$ and $\psi^\mu$ should ensure that both
 (\ref{fer-bc}) and (\ref{variation}) are satisfied. The bosonic
 boundary condition (\ref{bosbound})
will cancel the bosonic variation in (\ref{fer-bc}). 
 Using it in  (\ref{variation})
 and trying to choose the fermionic boundary conditions such
 that both (\ref{fer-bc}) and (\ref{variation}) are satisfied,
 however, we run into contradictions. 
 In the presence of the $B$-field there is no such fermionic 
 boundary condition\footnote{There is a trivial solution
 $\psi^\mu_{-}=\pm \psi^\mu_{+} = const.spinor$ which we are not
 intrested in.}. 

 Without changing the bosonic part,
 the way forward is to add boundary terms, i.e., total derivatives, to the
action (\ref{compaction}). 
For constant $B$-field, there are two (essentially unique) $2D$
Lorentz-invariant boundary terms involving two fermions and one derivative.
They give us the following additional action:
\be
\label{boundaryterm}
 S_{bound} = - \frac{1}{4\pi\alpha'} \int d^2\xi\,\left [
 \alpha \left(\epsilon^{\alpha \beta}B_{\mu \nu}\overline\psi^\mu
\rho_\alpha
\partial_\beta \psi^\nu\right )
 + \beta \left (
 B_{\mu \nu}\overline\psi^\mu \rho^\alpha \partial_\alpha \psi^\nu
\right)\right ].
\ee
 Analysing the sum of 
 (\ref{compaction}) and (\ref{boundaryterm}) we find solutions to the
relations corresponding to (\ref{fer-bc}) and (\ref{variation}) for 
$\alpha=\beta=i$. Thus the boundary term
 (\ref{boundaryterm}) has the form
\be
\label{bt}
 S_{bound} = -\frac{1}{4\pi\alpha'} \int d^2\xi\,(2i B_{\mu\nu}
 \psi_{+}^\mu \d_= \psi_{+}^\nu ) =  -\frac{1}{4\pi\alpha'} \int d^2\xi\,
 ( i\d_=(B_{\mu\nu}
 \psi_{+}^\mu  \psi_{+}^\nu) ),
\ee
 and the sum of the actions is
\be
\label{rightac}
S= -\frac{1}{4\pi \alpha'}\int d^2\xi
(\partial_\alpha X_\mu \partial^\alpha X^\mu
 + \epsilon^{\alpha \beta}B_{\mu \nu}\partial_\alpha X^\mu \partial_\beta
X^\nu
 +i E_{\nu\mu}
\overline\psi^\mu \rho^\alpha \partial_\alpha \psi^\nu).
\ee
 The boundary conditions are (\ref{bosbound}) and
\be
\label{ferb}
\left( E_{\nu\mu}\psi_{+}^\nu \mp
 E_{\mu\nu}\psi_{-}^\nu \right )|_{\sigma=0,\pi},
\ee
related as in (\ref{varfer}). 
When these conditions are imposed, the action (\ref{rightac}) is
supersymmetric. Clearly, this action cannot be written in standard
 bulk-superfield
 form due to the boundary term.

\subsection{Non-constant metric 
 and $B$-field}

In this subsection we extend the previous analysis to include a general
metric
$g_{\mu\nu}$ and antisymmetric
 two-form $B_{\mu\nu}$. For ease of notation, we use superfield language at
some of the steps where we used component notation in the previous
subsection. 

Again we start from the superfield version of the theory\footnote{In this
section
 we use the  normalization $\alpha'=(4\pi)^{-1}$.}
\beq
S = \int d^2 \xi d^2 \theta\,\, {\cal L}=
\int d^2 \xi d^2 \theta\,\,  D_+\P^\mu D_-\P^\nu E_{\mu\nu}(\Phi),
\eeq{lagr}
 where the $\theta$ independent part (denoted by $|$) of $E$ is
$E_{\mu\nu}|\equiv g_{\mu\nu}+B_{\mu\nu}$ with $g$ and $B$ the space-time
metric and antisymmetric tensor field, respectively.
 
 Using (\ref{tfs}) of the appendix, we find the supersymmetry variation of a
general
 action to be
\beq
\delta_\e S =
-i\int d^2\xi\left\{\e^+\d_\+D_--\e^-\d_=D_+\right\}{\cal L}|.
\eeq{dels}
For the special case of (\ref{lagr}), 
 (\ref{dels}) implies that
\beq
 \eta\left\{(
D_-\pm D_+)D_+\P^\mu D_-\P^\nu E_{\mu\nu}(\P )\right\}|
\eeq{susy}
has to vanish at $\s =0,\pi$. (Here $\eta \equiv \e^+=\pm \e^- $, as
before). 

The field equations for the action with Lagrangian (\ref{lagr}) are
obtained from the general variation
\ber
\delta S &=& \int d^2\xi d^2\th \left[\delta
\P\left\{(D_+D_-\P^\nu)(E_{[\nu\mu]}
+D_+\P^\r D_-\P^\nu(E_{\r[\nu\mu]}-E_{\mu\nu\r})\right\}\right]\cr
&&+\int d^2\xi d^2\th \left[ D_+\left\{\delta \P^\mu D_-\P^\nu
E_{\mu\nu}\right\}-
D_-\left\{D_+\P^\mu \delta \P^\nu E_{\mu\nu}\right\}\right]\cr
&&\equiv I_1+I_2.
\eer{feqs}
Here $I_1$ gives the field equations in the bulk (of the world sheet)
and $I_2$ is a boundary term which implies the vanishing of
\beq
\left[ D_-\left\{\delta \P^\mu D_-\P^\nu E_{\mu\nu}\right\}
-D_+\left\{D_+\P^\mu \delta \P^\nu E_{\mu\nu}\right\}\right] |
\eeq{bdy}
at $\s =0,\pi$.

As before, it is easy to convince oneself that the two requirements
(\ref{susy}) and
(\ref{bdy}) are incompatible except when $B_{\mu\nu}=0$.
Since the discrepancy is purely at the boundary of the world sheet we
want to add
a boundary term. Guided by a study of the case of constant
$E_{\mu\nu}$ (see equation (\ref{bt})), we
add the term
\beq
{\cal L}_B=-i\d_=(\p_+^\mu\p_+^\nu B_{\mu\nu}(X))
\eeq{bterm}
to the component Lagrange density. Adding the contributions from
(\ref{bterm}) to (\ref{bdy}), the boundary term (in components)
reads
\ber
 &-&\left\{i\left[\delta \p^\mu_-\p^\nu_--\delta
\p^\mu_+\p^\nu_+\right]E_{\mu\nu}\right.\cr &+&\left.\delta
X^\mu\left[\d_=X^\nu E_{\mu\nu}-\d_\+X^\nu E_{\nu\mu}
+i\p^\nu_-\p^\r_-E_{\mu\nu ,\r}-i\p^\nu_+\p^\r_+E_{\nu [\mu ,\r ]}
\right]\right\}_{\s =0,\pi}=0.
\eer{bdy2}
(Comma denotes a partial
derivative.) 
Our solution for the present case is obtained starting
from formally the same fermionic condition as in the constant case, i.e.,
from (\ref{ferb}) with non-constant $E_{\mu\nu}$. Substituting the $\p$
variations (at the boundary) that result from
(\ref{ferb}) into (\ref{bdy2}) we find that (\ref{bosbound}) gets replaced
by
\beq
\left[i\left\{ \d_\+X^\nu E_{\nu\mu}-\d_=X^\nu E_{\mu\nu}\right\}
\pm \p^\r _-\p^\s_+ E_{\s\r ,\mu}+\p^\s _-\p^\r_- E_{\mu\s ,\r}
-\p^\s _+\p^\r_+ E_{\s\mu ,\r}\right]_{\s
=0,\pi}=0
\eeq{xsol2}
The relations (\ref{ferb}) and (\ref{xsol2}) in conjunction with the
$F$-field equations are sufficient to show that the sum of the
boundary supersymmetry variation (\ref{susy}) and the variation of
(\ref{bterm})
vanish. The auxiliary $F$-field equation follows from
$I_1$ in (\ref{feqs}) and
reads
\be
\label{Feq}
2F^\nu_{+-} g_{\mu\nu}+\p_+^\r\p_-^\nu\left(E_{\mu\nu ,\r}+
 E_{\r\mu ,\nu}-E_{\r\nu ,\mu}\right)=0.
\ee
We note for future reference that it contains the $B$-field as a
field- strength only.
 One can check that the boundary supersymmetry variation of (\ref{ferb}) 
 is proportional to  (\ref{xsol2}), the same
 relation as in (\ref{varfer}). In checking this one should keep
 in mind that the supersymmetry transformations look rather involved
 due to (\ref{Feq}).

In the constant case 
 there are only two
 fermionic boundary terms  possible. However, in the 
 nonconstant case there are infinite many  fermionic boundary terms
available (for instance, $B_{\mu\nu ,\rho}\p_+^\mu \p_+^\nu
 \p_+^\rho$). All these terms (except the terms in 
 (\ref{boundaryterm})) contain derivatives of the background fields.

\section{Matter coupling on the boundary}

In this section we discuss the possible coupling to an $A_\mu$-field on the
boundary. We argue that there is an essentially unique such coupling that
preserves supersymmetry and the shift symmetry (defined below).

The bosonic sigma model with a $B$-field coupling is invariant up to
boundary terms under the transformation
\be
\label{shift}
\delta B_{\mu\nu}=\d_{[\mu}\L_{\nu]}.
\ee
When a boundary is present the resulting boundary term is compensated by a
shift $\delta A_\mu=\L_\mu$ of an $A$-field on the boundary whose action
is 
\be
\label{Aact}
S_A=\int d\tau A_\mu\dot{X}^\mu .
\ee 
When we consider the supersymmetric sigma model, two problems confront us:
The addition of an $A$-action will change the boundary conditions and we
must preserve supersymmetry. Remarkably, both these can be resolved. 

We start from the supersymmetric action of the previous section,
i.e., (\ref{lagr}) with the addition of (\ref{bterm}). We then observe that a
field redefinition 
$B_{\mu\nu}\to B_{\mu\nu}+F_{\mu\nu}\equiv \hat{B}_{\mu\nu}$ will only give a
contribution on the boundary, due to the invariance (\ref{shift}). We
collect all the $A$-terms to a boundary action
\be
\label{Aact2}
S_A=\int d\tau\left(-2 A_\mu\dot{X}^\mu +\ihalf
(\p_+^\mu +\p_-^\mu)F_{\mu\nu}(\p_+^\nu -\p_-^\nu)\right).
\ee 
This is not the usual supersymmetrization of (\ref{Aact}) 
 (see (\ref{interaction}) below); one has to keep
in mind that the supersymmetry only holds modulo the boundary conditions.
These now read
\be
\label{ferb2}
\left( \hat{E}_{\nu\mu}\psi_{+}^\nu \mp
 \hat{E}_{\mu\nu}\psi_{-}^\nu \right )|_{\sigma=0,\pi},
\ee
and
\beq
\left[i\left\{ \d_\+X^\nu \hat{E}_{\nu\mu}-\d_=X^\nu \hat{E}_{\mu\nu}\right\}
\pm \p^\r _-\p^\s_+ \hat{E}_{\s\r ,\mu}+\p^\s _-\p^\r_-
\hat{E}_{\mu\s ,\r} -\p^\s _+\p^\r_+ \hat{E}_{\s\mu ,\r}\right]_{\s
=0,\pi}=0,
\eeq{xsol22}
where $\hat{E}$ contains $B$ and $F$ in the combination $\hat{B}$.
  Note that this implies that the boundary
conditions are invariant under the shift symmetry. Note also that the
auxiliary $F^\mu_{+-}$ field equations that we need in the supersymmetry
check are invariant too (see the comment after (\ref{Feq})).
 In fact, since our present model is related to the
supersymmetric one by a shift invariant field redefinition, it is
supersymmetric and shift-invariant by construction. This has also been
explicitly verified.

Some comments are in order.
First, we again stress that the field-redefinition
of $B$ is a tool which allows us to identify the
shift-invariant action. Second, superficially, (\ref{Aact2})
depends on both
$\p_++\p_-$ and
$\p_+-\p_-$. However, using (\ref{ferb2}) we 
 may eliminate one in favour of the other.
Third, one
may ask what happens to our model in the limit of vanishing
$B$-field and the relation to the standard supersymmetric $A$-field action
as given in (\ref{interaction}) below. 
When $B=0$ there are still $F$-contributions to (\ref{ferb2}-\ref{xsol22}),
and they do indeed ensure supersymmetry (modulo these conditions). The usual
$A$-field coupling, on the other hand, will give contributions to the
boundary conditions that are in fact incompatible with supersymmety of the
full action.

\section{Covariant quantization}

 In this section we would for completeness like to sketch the covariant
 quantization
 of the model and specifically discuss the issue of the broken Lorentz 
 group. At the end the construction of the vertex operator for emission
 of massless boson is given and some problems related to the
 shift symmetry are pointed out.

 Let us look at the theory with constant $B_{\mu\nu}$ as given by
(\ref{rightac}), 
 with Neumann boundary conditions in all directions. 
 Since the equations of motion are not modified we can solve them in the
 standard way. The presence of a constant $B$-field results in the
 boundary conditions which relate the left and right
 movers in a non-trivial way \cite{Chu:1998qz}. 
 The general solution of the bosonic equation
of motion,
 satisfying the bosonic boundary condition, has the form
\be
\label{X}
X^\mu(\tau,\sigma)= q^\mu + 2\alpha' (G^{-1})^{\mu\nu} p_\nu \tau -
2\alpha' \theta^{\mu\nu} p_\nu \sigma  
+ \sqrt{2\alpha'} \sum\limits_{n \neq 0} \frac{e^{-in\tau}}{n} (
 i \alpha^\mu_n \cos n\s + B^\mu{_\nu} \alpha^\nu_n \sin n\s),  
\ee
where we use the notation  (for details see \cite{Fayyazuddin:1999gy})
\be
\label{notat}
 G_{\mu\nu} \equiv E_{\rho\mu} \eta^{\rho\sigma} E_{\sigma\nu} =
 \eta_{\mu\nu} - B_{\mu\rho} \eta^{\rho\sigma} B_{\sigma\nu},
\,\,\,\,\,\,\,\,\,\,\,\,\,\,\,
 \theta^{\mu\nu} \equiv - B^\mu{_\sigma}(G^{-1})^{\sigma\nu}.
\ee
 We should also solve the equations of motion for the
 fermionic coordinates
\be
\label{eqmot}
\d_{=}\psi^\nu_{+} =0,\,\,\,\,\,\,\,\,\,\,\,\,\,\,\,
\d_{\+}\psi^\nu_{-}=0,
\ee
taking into account the fermionic boundary conditions (\ref{ferb})
\be
\label{bfc}
  E_{\nu\mu} \psi^\nu_{+} \mp E_{\mu\nu}
 \psi^\nu_{-}|_{\sigma=0,\pi} =0,
\ee
 where the plus sign corresponds to Neveu-Schwarz (NS) and
 the minus to Ramond (R) conditions.
 As usual, the overall relative sign is conventional, so 
 without loss of generality we set  $E_{\nu\mu} \psi^\nu_{+}(\tau, 0)=
 E_{\mu\nu} \psi^\nu_{-}(\tau, 0)$.
  We thus have the following solutions of (\ref{eqmot}) and
 (\ref{bfc}):
\be
 \psi_{-}^\mu = \eta^{\mu\nu}  E_{\rho\nu} \frac{1}{\sqrt{2}}
\sum\limits_{r \in Z+1/2} b_r^\rho e^{ir(\tau
-\sigma)}
\ee
\be
\psi_{+}^\mu = \eta^{\mu\nu} E_{\nu\rho}
 \frac{1}{\sqrt{2}}
\sum\limits_{r \in Z+1/2} b_r^\rho e^{ir(\tau
+\sigma)}
\ee
 for the NS sector and
\be
\psi_{-}^\mu = \eta^{\mu\nu} E_{\rho\nu}
 \frac{1}{\sqrt{2}}
\sum\limits_{ n
\in Z} d_n^\rho e^{in(\tau
-\sigma)}
\ee
\be
\psi_{+}^\mu = \eta^{\mu\nu} E_{\nu\rho}
 \frac{1}{\sqrt{2}}
\sum\limits_{ n
\in Z} d_n^\rho e^{in(\tau +
\sigma)}
\ee
 for the R sector. The standard anticommutation relations for
 the fermionic coordinates
\be
\{\psi_{A}^\mu(\tau,\sigma),\psi_{B}^\nu(\tau, \sigma')\}
= \pi \eta^{\mu\nu} \delta(\sigma-\sigma) \delta_{AB}
\ee
 imply anticommutation relations for the modes
\be
\label{A1}
\{ b_r^\mu, b_s^\nu \} = (G^{-1})^{\mu\nu} \delta_{r+s},
\,\,\,\,\,\,\,\,\,\,\,\,\,\,\,\,\,\,
\{ d_n^\mu, d_m^\nu\} =  (G^{-1})^{\mu\nu} \delta_{n+m}
\ee
 The canonical commutation relations for the bosonic counterpart imply
 the following commutator relations 
\be
\label{A2}
 [\alpha_n^\mu, \alpha_m^\nu] = n\delta_{n+m} (G^{-1})^{\mu\nu},
\,\,\,\,\,\,\,\,\,\,\,\,
 [q^\mu, p_\nu] = i \delta^\mu_\nu,
\,\,\,\,\,\,\,\,\,\,\,\,
 [q^\mu, q^\nu]= 2\pi i\alpha' \theta^{\mu\nu}.
\ee
The super Virasoro generators are
\be
\label{A3}
L_n = \frac{1}{2} \sum\limits_{m=-\infty}^{\infty} \alpha_{-m}^\mu
 G_{\mu\nu} \alpha_{m+n}^\nu + \frac{1}{2}
\sum\limits_{r=-\infty}^{\infty}
 (r+ \frac{1}{2}n) b_{-r}^\mu G_{\mu\nu}
b_{n+r}^\nu\,\,\,\,\,\,\,\,\,\,\,\,
\,\,\,\,(NS)
\ee
\be
\label{A4}
L_n = \frac{1}{2} \sum\limits_{m=-\infty}^{\infty} \alpha_{-m}^\mu
 G_{\mu\nu} \alpha_{m+n}^\nu + \frac{1}{2}
\sum\limits_{m=-\infty}^{\infty}
 (m+ \frac{1}{2}n) d_{-m}^\mu G_{\mu\nu}
d_{m+n}^\nu\,\,\,\,\,\,\,\,\,\,\,\,
\,\,\,\,(R)
\ee
\be
\label{A5}
G_r = \sum\limits_{n=-\infty}^{\infty} \alpha_{-n}^\mu G_{\mu\nu}
 b_{r+n}^\nu\,\,\,\,\,\,\,\,\,\,\,\,
\,\,\,\,(NS)
\ee
\be
\label{A6}
 F_n = \sum\limits_{m=-\infty}^{\infty} \alpha_{-m}^\mu G_{\mu\nu}
 d_{n+m}^\nu\,\,\,\,\,\,\,\,\,\,\,\,
\,\,\,\,(R),
\ee
 where   normal ordering is assumed in all expressions. These generators
 give the standard super Virasoro algebra with central extention.
 The $B$-field does not change the anomaly in the super Virasoro algebra
 and the system can thus be quantized as usual.

 The presence of a $B$-field breaks the Lorentz symmetry to
 the $SO(2r-1,1)\otimes (SO(2))^{d/2-r}$ subgroup
 where $(d-2r)$ is the rank of the matrix $B_{\mu\nu}$. 
 Let us take a look at the spectrum. As in the bosonic case, 
 to avoid trouble we should
 define the mass using the new metric $G_{\mu\nu}$.
 The NS sector is the same as usual. The ground state corresponds
 to a tachyon with mass $M^2 = -
 p_\mu (G^{-1})^{\mu\nu} p_\nu = -1/(2\alpha')$. The state
 $\zeta_\mu b_{-1/2}^\mu |0,k\rangle$ is a  massless vector
 with respect to the new Lorentz group.
 In the R sector we have a fermionic zero mode
 that makes the R ground state degenerate, since $[d_0^\mu, L_0]=0$.
\be
\{ d_0^\mu, d_0^\nu \} = (G^{-1})^{\mu\nu}.
\ee
 Thus the R ground state transforms as a space-time fermion under 
 the new Lorentz group. The zero modes are given by 
\be
 d_0^\mu = \frac{1}{\sqrt{2}}
 \left (\frac{1}{\eta-B} \right)^{\mu\nu} \Gamma_\nu
\ee
 where $\Gamma_\nu$ are the standard 
 gamma matrices as in \cite{Green:1987sp}. One can interpret
this to mean  
 that as the $B$-field varies from $0$ to $\infty$
 there is smooth interpolation between Lorentz symmetry and the R-symmetry. 
 
 We may use the above results to construct the vertex operator
for 
 emission of massless boson $\zeta_\mu b_{-1/2}^\mu |0,k\rangle$ 
 along the 
 standard lines (see discussion in section 4.2.3 i volume 1 
 of \cite{Green:1987sp})
 using the modified commutation relations (\ref{A1}),
(\ref{A2}) 
 and the super Virasoro generators (\ref{A3})-(\ref{A6}). The
 result is 
\be
\label{vertop}
 V= (\zeta_\mu \dot{X}^\mu(0) - \zeta_\mu \Psi^\mu(0) k_\nu
 \Psi^\nu(0) ) e^{ik_\mu X^\mu(0)}
\ee  
 where $\Psi^\mu(0)=1/2(\psi^\mu_{+}(0)+ \psi^\mu_{-}(0))$  and
$\zeta_\mu$ is the polarization vector of the spin-one field.
Note that (\ref{vertop}) is identical to that discussed in
\cite{Seiberg:1999vs}. If we naively read off the matter
coupling to the sigma model \cite{Callan:1988wz}, we expect it to be 
\be
\label{interaction}
S_A = \int d\tau \left (A_\mu \dot{X}^\mu - \frac{1}{2i} F_{\mu\nu} 
 \Psi^\mu \Psi^\nu \right ).
\ee
 This concides with the boundary interactions discussed in 
 \cite{Seiberg:1999vs, Hashimoto:1999dq, Andreev:2000pv} 
 but disagrees with (\ref{Aact2}). 
 The difference seems to emanate from our different
 approaches. The interaction (\ref{interaction}) is
 supersymmetric by itself, independent of the boundary conditions. 
 The interaction (\ref{Aact2}) on the other hand, is only
supersymmetric together with the rest of the 
 action and  with the appropriate boundary conditions
(derived from the total action) imposed. Note also that the full
action with
 (\ref{Aact2}) included respects the shift symmetry 
 whereas we do not know how to realize the shift symmetry in 
 a supersymmetric way with the interaction in
 (\ref{interaction}).

\section{Discussion}

We first make some comment on the boundary conditions that we have derived.
Let us start from the case when
$B_{\mu\nu}=0$
 and $g_{\mu\nu}$ is arbitrary. In this situation the  boundary
 conditions (\ref{ferb}) for the fermions are
\be
\label{bcwg}
 \{\p_+^\mu \mp \p_-^\mu\}_{\s =0,\pi} =0.
\ee
 The F-field equation (\ref{Feq}) has the form 
\be
\label{Feqwg}
 F^\mu_{+-} +\p_+^\r\p_-^\nu \Gamma^{\mu}{_{\nu\rho}} =0,
\ee
 where $\Gamma^{\mu}{_{\nu\rho}}$ are the Christoffel symbols. Thus on the
 boundary the equation (\ref{Feqwg}) reduces to $F^\mu_{+-}=0$ 
 because of the symmetries of the Christoffel symbols and the boundary 
 conditions (\ref{bcwg}). Therefore the supersymmetric transformation
 restricted to the boundary is exactly the same as in the 
 constant case (\ref{susytransf}). Further, the boundary 
 condition (\ref{xsol2}) collapses to  $X'^\mu =0$.
 We thus see that the curved metric by itself 
  does not make the boundary conditions more complicated than 
 in the constant case. However some problems might arise when we
 try to introduce  Dirichlet conditions in some of the directions. 
 Like in the case with constant $B$-field, discussed below, the mixed
components
 of the metric $g_{im}$  are the source of these difficulties.
 (Here $i$ is Dirichlet and $m$ is Neumann 
 directions.)

In our discussion so far we have used Neumann bounary 
 conditions\footnote{By Neumann boundary conditions we mean conditions 
 that ensure that
 there is no flow of momentum off the endpoints of the string, 
 not to be confused with the mathematical 
 notion of Neumann boundary conditions.} in all directions. 
 Thus we had in mind, e.g., open strings that represent
fluctuations of a D9-brane.
  Restricting to  constant $E$ case, we may ask which other
  boundary conditions the action
(\ref{rightac}) admits.  
  In other words we take a look at a general 
 Dp-brane with $p<9$. We assume no special direction for the background
$B_{\mu\nu}$-field\footnote{One may view this as
 embedding a Dp-brane in a D9-brane. We shall see that in 
 the presence of a $B$-field this cannot be  arbitrarily done.}. 
 To define the Dp-brane we impose the following Neumann conditions
\be
\label{Nbc}
\d_= X^n E_{mn} - \d_\+ X^n E_{nm}|_{\sigma=0,\pi} =0,
\,\,\,\,\,\,\,\,\,\,\, E_{nm}\psi_{+}^n \mp
 E_{mn}\psi_{-}^n|_{\sigma=0,\pi}=0,\,\,\,\,\,\,\,\,\,\,\,
n,m=0,1,...,p\,\,\,
\ee   
 and the Dirichlet conditions
\be
\label{Dbc}
\dot{X}^i=0|_{\sigma=0,\pi},\,\,\,\,\,\,\,\,\,\,\,\psi_{+}^i \pm
 \psi_{-}^i|_{\sigma=0,\pi}=0,\,\,\,\,\,\,\,\,\,\,\,i=p+1,...,9.
\ee
 In (\ref{Nbc}) and (\ref{Dbc}) the bosonic and fermionic conditions 
 are related to each other through the supersymmetry transformations
 (\ref{susytransf}). By plugging these conditions into the corresponding 
 variations of the action (\ref{rightac}) one will find that
 in the generic situation
 there are non-vanishing terms 
 propotional to the mixed component $B_{im}$. 

 One may view this slightly differently: If we impose the Dirichlet
 conditions (\ref{Dbc}) in some directions and try to choose other
 boundary conditions to cancel the corresponding variations
 then in addition to (\ref{Nbc}) and (\ref{Dbc}) we would have to introduce 
  boundary conditions which mix the Neumann and Dirichlet
 directions. A theory with such  
 boundary conditions would be inconsistient, as may be easily seen. 
  However, by a meticulous
 choice of the Dirichlet directions we may obtain that
  $B_{im}=0$ and then things will
 work out. We may thus interpret this as restrictions on the orientation
 of those Dp-branes for which the action (\ref{rightac}) is consistent. 
 E.g., there is no problem if the  $B$-field is
non-zero
 only along the brane. The restrictions have a natural interpretation 
 related to the broken Lorentz group.

\begin{flushleft} {\Large\bf Acknowledgments} \end{flushleft}

 We are grateful to Ansar Fayyazuddin, Subir Mukhopadhyay, 
 Martin Ro\v{c}ek and Rikard
von Unge for comments.

\appendix
\section{Appendix}

Throughout the paper we use $\mu,\nu,...$ as spacetime indices.
The two dimensional spinor indices
are $(a,b,...=0,1)$ and $(\alpha,\beta,...=+,-)$ denote world sheet
indices.
We also use superspace conventions where the spinor coordinates are
labeled $\th^{\pm}$
and the covariant derivatives $D_\pm$ and supersymmetry generators
$Q_\pm$ satisfy
\ber
D^2_+ &=&i\d_\+, \quad
D^2_- =i\d_= \quad \{D_+,D_-\}=0\cr
Q_\pm &=& -D_\pm+2i\th^{\pm}\d_{\pp}
\eer{alg}
 where $\d_{\pp}=\partial_0\pm\partial_1$.
In terms of the covariant derivatives, a supersymmetry transformation of
a
superfield
$\P$ is then given by
\ber
\delta \P &\equiv & (\e^+Q_++\e^-Q_-)\P \cr
&=& -(\e^+D_++\e^-D_-)\P
+2i(\e^+\th^+\d_\++\e^-\th^-\d_=)\P
\eer{tfs}
The components of a superfield $\P$ are defined via projections as
follows:
\ber
\P|\equiv X, \quad D_\pm\P| \equiv \p_\pm, \quad D_+D_-\P|\equiv F_{+-}
,
\eer{comp}
where a vertical bar denotes ``the $\th =0$ part of ''.
Choosing the world-sheet $\gamma$-matrices as
\bea
\rho^{0}= \left(\begin{array}{clcr}
0 & -1 \\
1 & 0
\end{array}\right),\;
\rho^{1}= \left(\begin{array}{clcr}
0 & 1 \\
1 & 0
\end{array}\right),\;
\rho^{3}= \left(\begin{array}{clcr}
-1 & 0 \\
0 & 1
\end{array}\right),
\eea
where $\{\rho^\alpha, \rho^\beta\} = +2\eta^{\alpha\beta}$ and
$\eta^{\alpha\beta}=(-,+)$,
the Majorana spinors $\p$ can be decomposed into two compenents with
 different chirality
\be
 \psi^{\mp} = \frac{1\pm \rho^3}{2} \psi,\,\,\,\,\,\,\,\,\,\,\,\,\,\,
 \rho^3 = \rho^0\rho^1.
\ee
 Thus the spinors $\p$  are
\be
\psi = \left(\begin{array}{l}
            \psi^{+}\\
            \psi^{-}
            \end{array} \right)
\ee
 and $\bar{\p}\equiv(\p_{+}, \p_{-})=\p^{\dagger} \rho^{0}$.

\newpage
\noindent



\begin{thebibliography}{99}
%
\bibitem{Connes:1998cr}
A.~Connes, M.~R.~Douglas and A.~Schwarz,
``Noncommutative geometry and matrix theory: Compactification on tori,''
JHEP {\bf 9802} (1998) 003
[hep-th/9711162].
%
\bibitem{Chu:1998qz}
C.~Chu and P.~Ho,
 ``Noncommutative open string and D-brane,''
Nucl.\ Phys.\ {\bf B550} (1999) 151
hep-th/9812219.
%
\bibitem{Seiberg:1999vs}
N.~Seiberg and E.~Witten,
 ``String theory and noncommutative geometry,''
JHEP {\bf 09} (1999) 032
hep-th/9908142.
%
\bibitem{Schomerus:1999ug}
V.~Schomerus,
 ``D-branes and deformation quantization,''
JHEP {\bf 06} (1999) 030
hep-th/9903205.
%
\bibitem{Chu:1999gi}
C.~Chu and P.~Ho,
``Constrained quantization of open string in background B
  field and  noncommutative D-brane,''
hep-th/9906192.
%
%
\bibitem{Fayyazuddin:1999gy}
A.~Fayyazuddin and M.~Zabzine,
``A note on bosonic open strings in constant B field,''
 to appear in Phys.Rev.D [hep-th/9911018].
%
\bibitem{Green:1987sp}
M.~B.~Green, J.~H.~Schwarz and E.~Witten,
``Superstring Theory. Vol. 1: Introduction,''
{\it  Cambridge, UK: Univ. Press. ( 1987) 469 P.
 ( Cambridge Monographs On Mathematical Physics)}.
%
\bibitem{Callan:1988wz}
C.~G.~Callan, C.~Lovelace, C.~R.~Nappi and S.~A.~Yost,
``Loop Corrections To Superstring Equations Of Motion,''
Nucl.\ Phys.\  {\bf B308} (1988) 221.
%
\bibitem{Hashimoto:1999dq}
K.~Hashimoto,
``Generalized supersymmetric boundary state,''
hep-th/9909095.
%
\bibitem{Andreev:2000pv}
O.~Andreev and H.~Dorn,
``On open string sigma-model and noncommutative gauge fields,''
Phys.\ Lett.\  {\bf B476} (2000) 402
[hep-th/9912070].
%

\end{thebibliography}
\end{document}